\documentclass[%
 reprint,
superscriptaddress,
 amsmath,amssymb,
 aps,
]{revtex4-1}

\usepackage{graphicx}
\usepackage{dcolumn}
\usepackage{bm}
\usepackage{xcolor}
\usepackage{ulem}

\newcommand{\SKIP}[1]{ }

\makeatletter
\newcommand{\ccaption}[2][]{%
	\begingroup
	\renewcommand{\@caption@fignum@sep}{ (color online). }%
	\caption[#1]{#2}%
	\endgroup%
}
\makeatother

\begin{document}

\preprint{APS/123-QED}


\title{Formation of transfermium elements in reactions with $^{208}$Pb}

\author{M. Albertsson}
\affiliation{Nuclear Science Division, Lawrence Berkeley National Laboratory, Berkeley, California 94720, USA}
\affiliation{Division of Mathematical Physics, Lund University, 221\,00 Lund, Sweden}
\author{B.G. Carlsson}
\affiliation{Division of Mathematical Physics, Lund University, 221\,00 Lund, Sweden}
\author{T. D{\o}ssing}
\affiliation{Niels Bohr Institute, University of Copenhagen, 2100 Copenhagen {\O}, Denmark}
\author{J. Randrup}
\affiliation{Nuclear Science Division, Lawrence Berkeley National Laboratory, Berkeley, California 94720, USA}
\author{D. Rudolph}
\affiliation{Division of Particle and Nuclear Physics, Lund University, 221\,00 Lund, Sweden}
\author{S. {\AA}berg}
\affiliation{Division of Mathematical Physics, Lund University, 221\,00 Lund, Sweden}

\date{\today}

\begin{abstract}
Within the Langevin framework,
we investigate the dynamics of the fusion process 
for production of transfermium elements in reactions of
$^{48}$Ca, $^{50}$Ti, $^{54}$Cr, and $^{58}$Fe with $^{208}$Pb.
After the reacting nuclei have made contact,
the early dynamical stage is dominated by 
the dissipation of the initial radial kinetic energy,
while the subsequent shape evolution is diffusive.
The probability for surmounting the inner barrier and forming a compound system
is obtained by simulating the evolution as a Metropolis random walk
in a five-dimensional potential-energy landscape.
Good agreement with the available data is obtained,
especially for the maximal formation probability.
\end{abstract}

\maketitle
\raggedbottom

\section{Introduction}\label{sec:intro}

The primary method for producing
superheavy atomic nuclei (SHN)
has been heavy-ion fusion reactions
\cite{dullman15:a,rudolph16:a}.
When relatively light reaction partners reach their Coulomb barrier
the configuration is more compact than that of the fission saddle shape.
Therefore,
a compound nucleus is automatically formed.
However, for heavier reaction partners
the Coulomb barrier is situated outside the fission saddle.
To avoid
a re-separation, referred to as {\it quasi-fission} (QF),
the shape of the composite system must diffuse over the inner barrier
\cite{swiatecki81:a,swiatecki82:a,bjornholm82:a}.

Thus the production of a
SHN in a fusion reaction
requires a heavy reaction system
and the process can be divided into three stages:
(i) {\it Contact:} 
the two reaction partners overcome their mutual Coulomb repulsion 
and achieve contact;
(ii) {\it Formation:} 
evolving in competition with QF, the combined system achieves 
a compact shape well inside the fission barrier;
(iii) {\it Survival:} the compact system deexcites towards the ground state
in competition with fission.

The heaviest known elements are produced in
fusion with targets
 heavier than $^{208}$Pb at rather large excitation energies.
However,
fusion reactions with $^{208}$Pb
are more accessible experimentally and thus
serve as a good starting point for theoretical studies
and we shall focus this first study on reactions
with the projectiles $^{48}$Ca, $^{50}$Ti, $^{54}$Cr, and $^{58}$Fe
on a $^{208}$Pb target.
For
three of these reactions, the formation probability
was obtained experimentally by extracting the fusion-fission component
from the total cross section \cite{banerjee19:a}.

The dynamics in the formation stage is not yet well understood and several models have been developed to describe the process. They can broadly be classified into two groups: time-dependent Hartree-Fock methods \cite{godbey20:a} and models based on the stochastic Langevin framework \cite{swiatecki03:a,swiatecki05:a,zagrebaev15:a}.

In this study we investigate the formation process with a Langevin-type treatment.
During the formation process, the nuclear shape is described by
five parameters and the associated
five-dimensional (5D)
potential-energy surface
includes energy-dependent shell and pairing effects.
The shape evolution is simulated in a Monte-Carlo manner
so an ensemble of events is generated,
thus allowing the extraction of both fluctuations of the observables
and correlations between them.

After the reacting nuclei have made contact,
the early part of the formation stage is assumed to be dominated
by the dissipation of the initial radial kinetic energy,
leading the shape along an effectively one-dimensional path
towards smaller overall elongations.
With the same assumption of overdamped shape evolution as invoked
in the fusion-by-diffusion (FBD) model \cite{swiatecki03:a,swiatecki05:a},
the subsequent shape evolution is then diffusion dominated.
A preliminary version of this treatment was published
in Ref.\ \cite{albertsson_thesis}.

Section \ref{sec:method_epot} describes
the shape coordinates employed to represent the composite system 
and the relevant structures in the potential-energy landscape, 
such as the different valleys.
Section
\ref{sec:dynamics} discusses the dynamics of the formation process
and the calculated formation probabilities are presented 
in Sec.\ \ref{sec:results}.
Finally, Sec.\ \ref{sec:summary}
provides
a summary and a discussion.

\section{Shapes and potential energy}
\label{sec:method_epot}

\begin{figure*}[bt]
\centering
\flushleft
\includegraphics[width=0.92\linewidth]{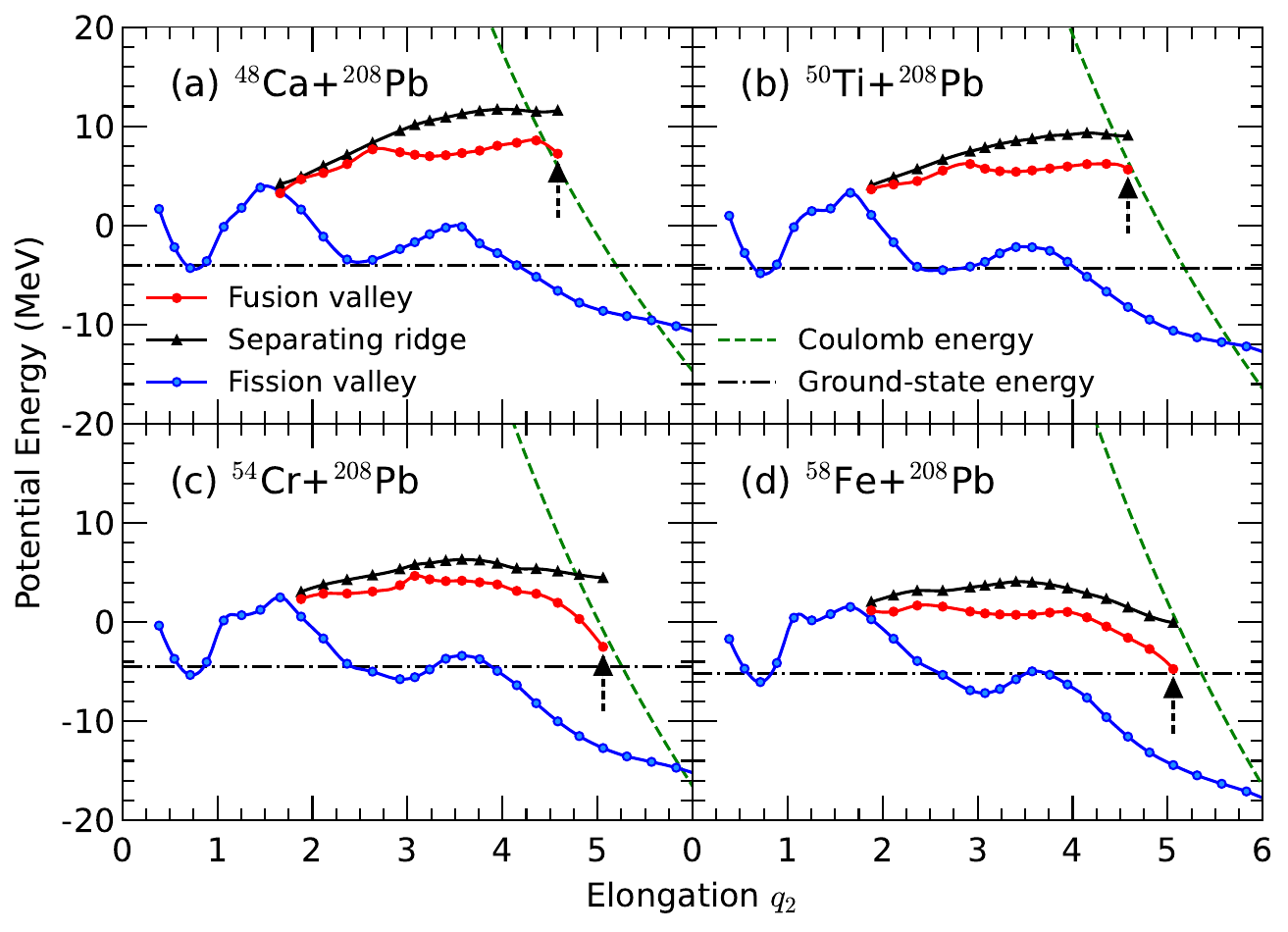}
\caption{\label{fig:epot1d}
 Potential energy for angular momentum $I$=0 as a function of elongation $q_2$ for reactions of the projectiles (a) $^{48}$Ca, (b) $^{50}$Ti, (c) $^{54}$Cr,
and (d) $^{58}$Fe with the target $^{208}$Pb.
 Blue curves correspond to the minimum potential energy for the mainly symmetric fission valleys,
 while red curves correspond to the asymmetric fusion valleys.
The curves connect the values at the shape lattice sites (indicated by solid
circles).
 The ridge between the two valleys is shown as
 black
 curves.
 The ground-state energy (located slightly above the first minimum due to the zero-point energy) is shown by the horizontal line.
The arrows indicate the contact elongations $q_2^{\rm cont}$ =
4.6, 4.6, 5.1 and 5.1, respectively.
The Coulomb potential between the corresponding two spheres 
are shown as dashed green lines;
these achieve touching at $q_2=2.9$, $3.0$, $3.2$, and $3.3$, respectively.
}
\end{figure*}

\begin{figure}[t]
 \begin{center}
 \includegraphics[width=1\linewidth]{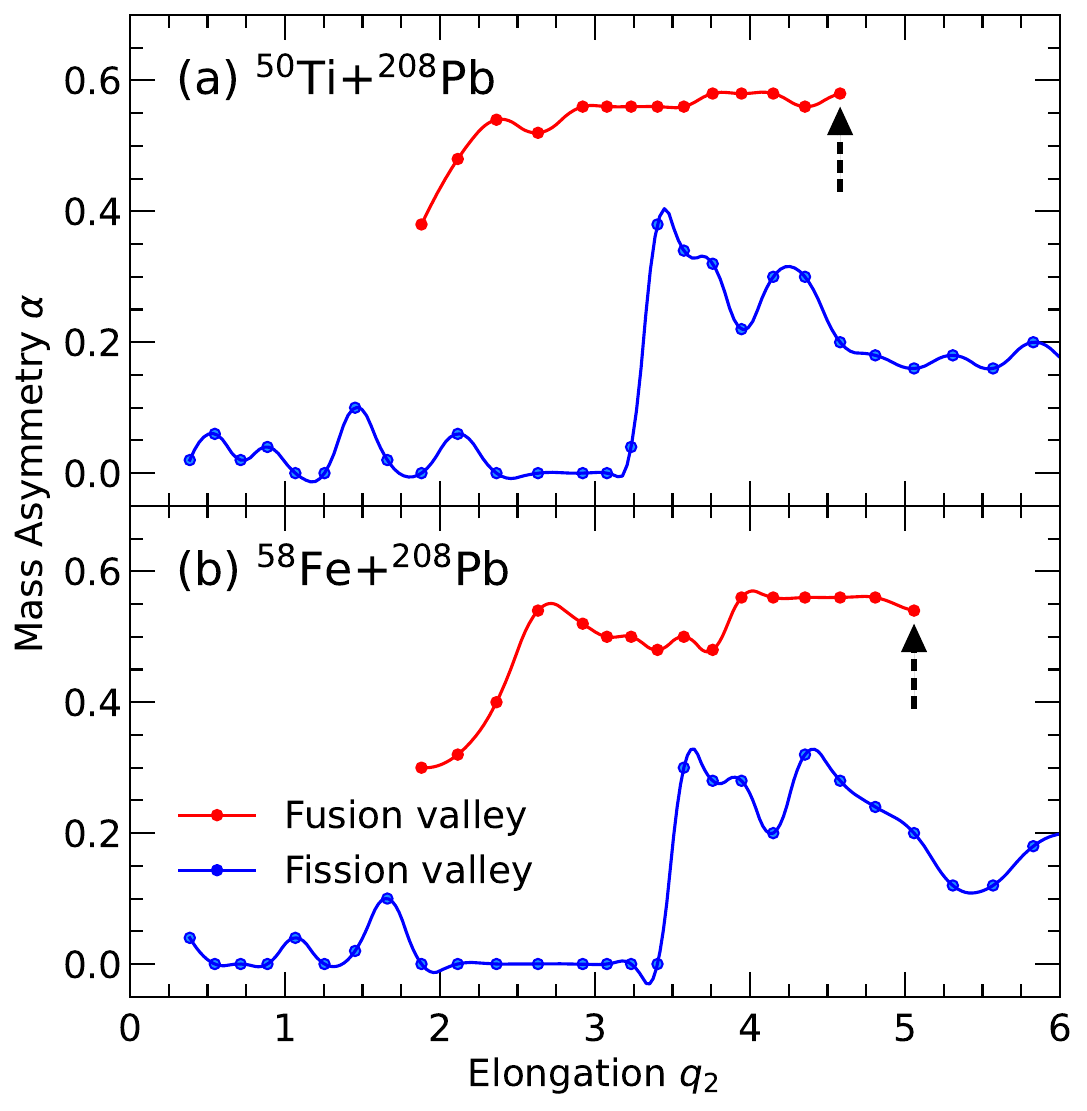}
 \caption{
 Mass asymmetry $\alpha$ as a function of the elongation $q_2$
 along the fission valley (blue) and the fusion valley (red)
for the reactions (a) $^{50}$Ti+$^{208}$Pb and (b) $^{58}$Fe+$^{208}$Pb.
The arrows indicate the mass asymmetry at contact.
 }
\label{fig:alpha}
 \end{center}
\end{figure}

\begin{figure}[bt]
 \begin{center}
 \includegraphics[width=1.0\linewidth]{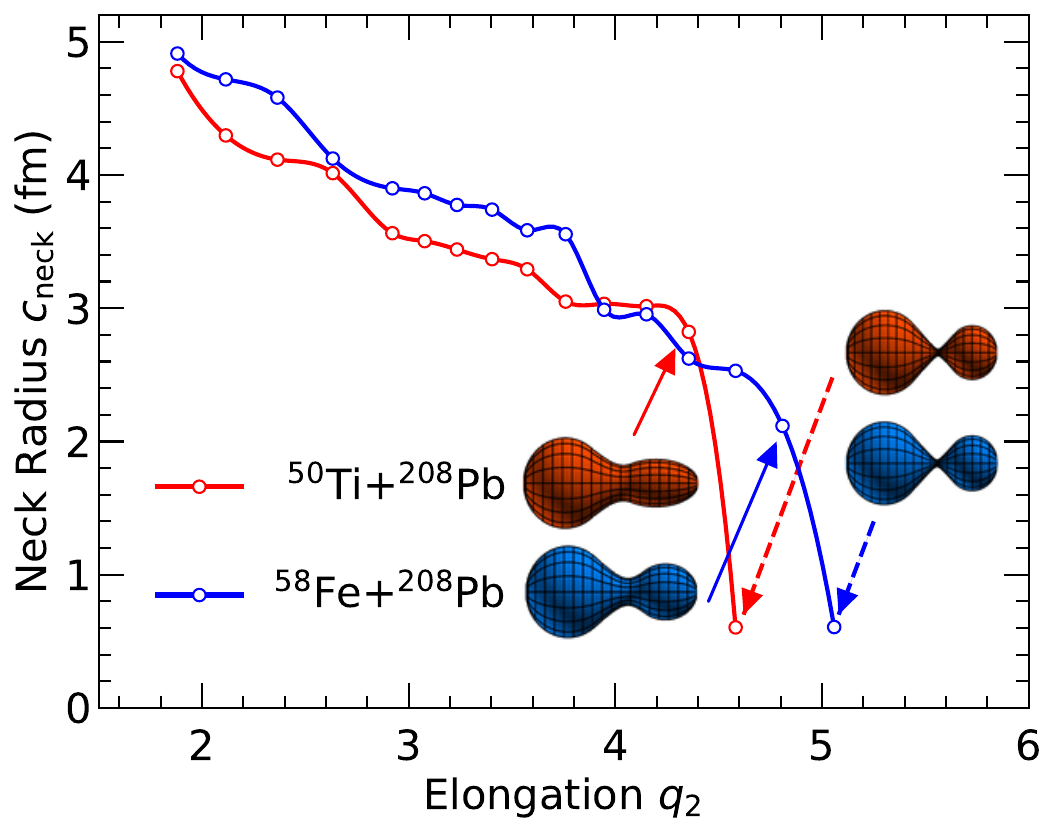}
 \caption{
Neck radius $c_{\rm neck}$ as a function of elongation $q_2$
along the fusion valley for 
$^{50}$Ti+$^{208}$Pb (red) and $^{58}$Fe+$^{208}$Pb (blue).
The shapes of the systems at contact (dashed arrows) 
and after contact (solid arrows) are also shown.
 }
\label{fig:neck}
 \end{center}
\end{figure}

The shape of the evolving nuclear system is described within
the three-quadratic-surface parametrization in which two spheroidal end parts 
are connected smoothly by a hyperbolic surface of revolution \cite{nix69:a}.
This shape family has five parameters: 
the overall elongation measured by the dimensionless quadrupole moment, $q_2$; 
the degree of reflection asymmetry, $\alpha$; 
the radius of the neck connecting the two parts, $c_{\rm neck}$; 
the deformations of the two spheroidal end parts,
$\epsilon_{\rm P}$ and $\epsilon_{\rm T}$. 
Collectively, these five shape parameters are denoted by $\boldsymbol{\chi}$.
The parameter $q_2$ is defined from the quadrupole moment of a 
uniformly charged sharp-surface volume, $Q_2$, 
as $q_2\equiv(4\pi Q_2)/(3ZR_A^2)$ with $R_A=1.2A^{1/3}$ fm.

For each shape, the potential energy $U(\boldsymbol{\chi})$ 
is calculated by the macroscopic-microscopic method
using the finite-range liquid-drop model \cite{moller09:a},
assuming that the system has a uniform charge-to-mass ratio.
The associated 5D shape lattice
is set up as
described
in Ref.\ \cite{albertsson20:b}.

The resulting potential-energy profile $U_{\rm min}(q_2)$,
{\it i.e.} the minimum energy of $U$ for a specified $q_2$,
is shown in Fig.\ \ref{fig:epot1d} for 
the reactions at angular momentum $I$=0 induced by the projectiles 
$^{48}$Ca, $^{50}$Ti, $^{54}$Cr, and $^{58}$Fe on a $^{208}$Pb target,
leading to the compound nuclei $^{256}$No (a), $^{258}$Rf (b), 
$^{262}$Sg (c), and $^{266}$Hs (d), respectively.

In all four cases the compound nucleus has
a prolate ground-state shape with $q_2\approx0.7$, corresponding to 
a quadrupole deformation of the overall shape of $\varepsilon_2\approx0.21$.
The inner saddle point protecting the compound nucleus from fissioning
is located at $q_2\approx1.5$,
as determined by the immersion method \cite{moller09:a}.
The saddle point is almost symmetric in all cases: $\alpha$=0.02 for projectiles $^{48}$Ca, $^{50}$Ti and $^{54}$Cr, and 
$\alpha$=0.1 for $^{58}$Fe.
Different valleys are found by identifying local minima for each $q_2$ value in the energy surface and smoothly following these minima as $q_2$ increases.
The ground states are connected to valleys that are mainly symmetric 
and these are denoted as the respective fission valleys (blue circles).
Being initially reflection symmetric,
the fission valleys bifurcate into asymmetric shapes at $q_2\approx3.5$
(see Fig.\ \ref{fig:alpha}).

For each case, in addition to the fission valley,
there is a valley (denoted as the fusion valley)
having a mass asymmetry similar to that of
the projectile-target system (red circles).
The ridge separating the two valleys is found using the immersion method
and is shown as a black curve. 
The shapes along the fusion valley are highly asymmetric with 
$\alpha \approx$ 0.5\textendash0.6 for $q_2\approx2.5$
but
become
less asymmetric for smaller $q_2$ (see Fig.\ \ref{fig:alpha}).

During the approach stage, while still well separated, the nuclei interact only via the
Coulomb potential,
which is shown for two spheres in Fig.\ \ref{fig:epot1d} as dashed green curves.
However, when the nuclei come close,
the nuclear interaction will start to have effect
and it is energetically more favourable to develop a neck.
This occurs at the {\it contact elongation} $q_2^{\text{cont}}$
(shown as black arrows in Fig.\ \ref{fig:epot1d}).

The neck radius along the fusion path is
shown in Fig.\ \ref{fig:neck}, where the
neck is seen to develop at
$q_2$=4.6 and 5.1
for the reactions  $^{50}$Ti+$^{208}$Pb and $^{58}$Fe+$^{208}$Pb,
respectively.
This rapid increase in neck radius is also in agreement with the ``neck zip'' concept discussed in the FBD model \cite{swiatecki03:a,swiatecki05:a}.
The nuclear interaction leads to a substantial lowering of the potential energy.
The shapes at contact and shortly after are also displayed 
in Fig.\ \ref{fig:neck} and it can be seen how the neck formation
causes a dramatic shape change of the projectile-like part,
from being nearly spherical to having a large prolate deformation.
Similar changes of the elongation of the projectile was found 
in the investigation of
fusion barriers in Ref.\ \cite{ichikawa05:a}.

\begin{figure}[t]
 \begin{center}
 \includegraphics[width=1\linewidth]{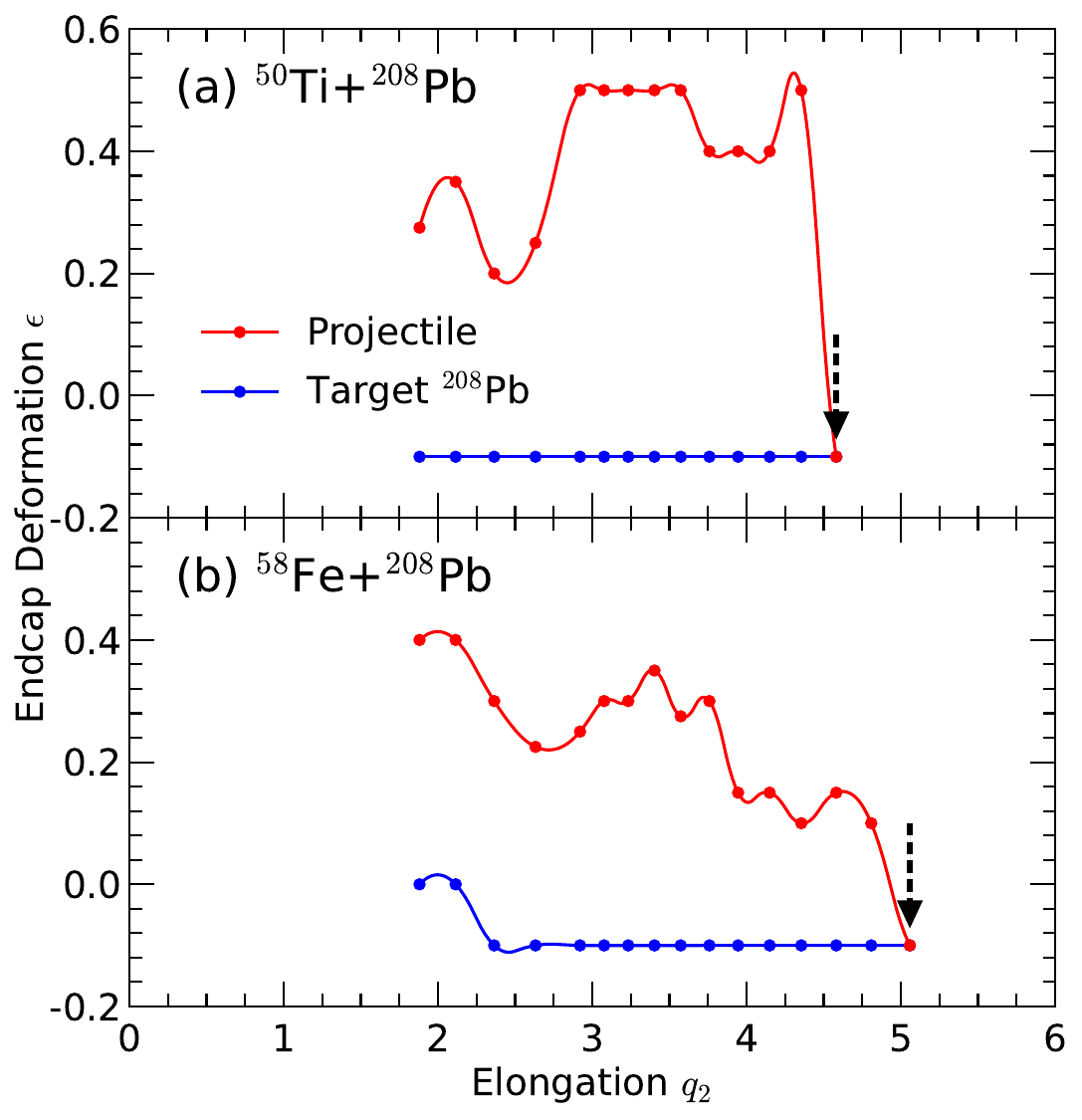}
 \caption{
Endcap quadrupole deformations
of the projectile, $\epsilon_{\rm P}$  (red)
 and the target, $\epsilon_{\rm T}$ (blue)
as functions of the overall elongation $q_2$ along the fusion valley
for the reactions (a) $^{50}$Ti+$^{208}$Pb and (b) $^{58}$Fe+$^{208}$Pb.
The arrows indicate the elongation of the projectile-target system at contact.
 }
\label{fig:eps}
 \end{center}
\end{figure}

The development of the optimal shapes along the fusion path 
is illustrated in Fig.\ \ref{fig:eps}.
While the target-like part retains a constant and slightly oblate shape,
the projectile-like part acquires a substantial and variable prolate shape,
with that of the titanium-like part building up quickly
and that of the iron-like part deforming more gradually.

The numerical calculations have been performed for different values of 
excitation energy $E_{\rm CN}^*$ and
angular momentum $I$.
The excitation energy of the compound nucleus is given by
\begin{equation}
E_{\rm CN}^{\ast} = E_{\rm cm} + Q,
\end{equation}
where $E_{\rm cm}$ is the relative kinetic energy
and 
\begin{equation}
Q=M_{\rm P}c^2+M_{\rm T}c^2-M_{\rm CN}c^2,
\end{equation}
with $M_{\rm P}$, $M_{\rm T}$, and $M_{\rm CN}$ 
being the masses of
projectile nucleus,
target nucleus,
and
corresponding compound nucleus, respectively.
The masses are calculated within the same macroscopic-microscopic model 
that is used to obtain the potential-energy surfaces \cite{moller04:a}.

Generally, the resulting compound nucleus inherits the angular momentum $I$
brought in by the colliding projectile-target system.
To a good approximation, this can be taken into account 
by adding a centrifugal potential, leading to an effective potential
\begin{equation}
U_I(\boldsymbol{\chi})=U(\boldsymbol{\chi})+\frac{\hbar^2I(I+1)}{2\mathcal{J}_\perp(\boldsymbol{\chi})},
\end{equation}
where
$\mathcal{J}_\perp(\boldsymbol{\chi})$ is the rigid-body moment of inertia perpendicular to the symmetry axis.
The rotational term corresponds to 
an increase of the potential energy at smaller
$q_2$ relative to larger $q_2$. 
For example, for $I$=60$\,\hbar$ in $^{50}$Ti+$^{208}$Pb, 
the centrifugal potential is 5.8\,MeV at contact,
while it is 9.9\,MeV at the fission saddle.

\section{Dynamics}
\label{sec:dynamics}

After establishing contact,
the two colliding nuclei continue their inward relative motion
while experiencing a frictional force that gradually converts
the relative kinetic energy into intrinsic excitation.
After this early drift-dominated stage 
has brought the relative motion to a halt,
the further shape evolution takes on a diffusive character 
in the full 5D shape space.
The system may then eventually become as compact as the ground-state shape
inside the inner barrier
(which we shall denote as fusion)
or
it may re-divide into two fragments (which is denoted as QF).

Section \ref{sec:method_contact} describes the contact configurations.
Then the drift stage is discussed in Sec.\ \ref{sec:method_drift}
and the diffusive stage is presented in Sec.\ \ref{sec:method_diffusion}. 

\subsection{Contact}
\label{sec:method_contact}

In order for the nuclei to come into contact, 
they first have to overcome the Coulomb barrier in the two-body channel. 
On the approach towards contact, vibrations and transfer channels are activated.
These processes effectively lead to a distribution in barrier heights, 
described by an average barrier height $B_0$ and a width $w$. 
Adopting this procedure, with the parameters derived from experiment
and discussed in Refs.\ \cite{siwek04:a,cap11:a},
we calculate the capture cross section as a function of energy,
$\sigma_{\rm capt}(E_{\rm CN}^*)$. 
Following Ref.\ \cite{cap11:a},
we use a sharp cut-off in angular momentum, 
assuming full transmission up to $I = I_{\rm max}$,
$T(I\leq I_{\rm max}) = 1$, and zero above, $T(I > I_{\rm max}) = 0$,
\begin{equation}
\label{eq:sigma_cap_1}
\sigma_{\rm capt} = \pi \lambdabar^2\sum_{I=0}^{\infty}(2I+1)T(I)
=\pi \lambdabar^2(I_{\rm max}+1)^2,
\end{equation}
where $\lambdabar$ is the reduced wavelength,
$\lambdabar^2=\hbar^2/2 \mu E_{\rm cm}$.

\begin{figure}[t]
 \begin{center}
 \includegraphics[width=1.0\linewidth]{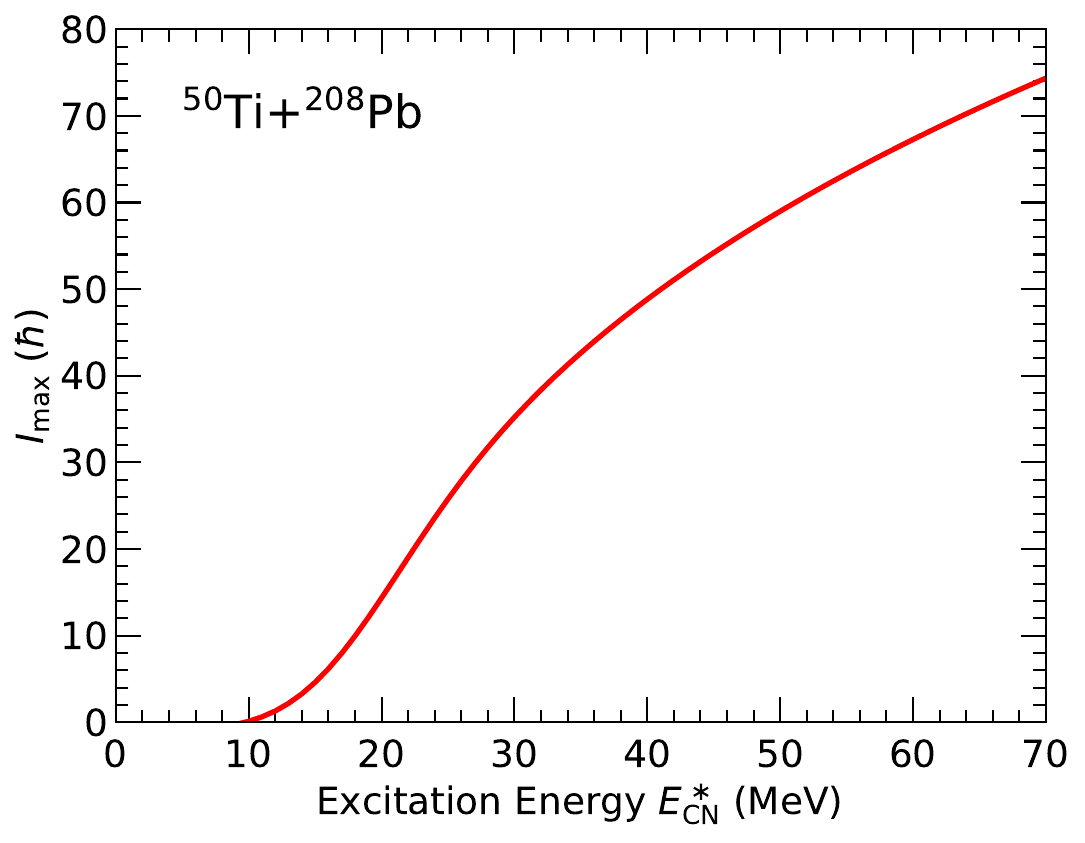}
 \caption{
Maximal angular momentum $I_{\rm max}$ as a function of
the excitation energy $E^*_{\rm CN}$ for $^{50}$Ti+$^{208}$Pb.
 }
\label{fig:Imax}
 \end{center}
\end{figure}

Figure \ref{fig:Imax} shows the resulting maximum angular momentum 
as a function of the excitation energy of the compound nucleus 
for $^{50}$Ti+$^{208}$Pb.
Its monotonic increase is steeper
for energies near the average Coulomb barrier height
$B_0=20$ MeV.

%
\subsection{Drift dynamics}
\label{sec:method_drift}

At contact the neck quickly develops and the radial friction sets in.
The inward motion then slows down as the associated kinetic energy 
is being dissipated into intrinsic excitation energy.
We assume that the system drifts in the radial
direction along the fusion valley without exploring other shape changes.
For each value of $q_2$ the lattice site with the lowest energy is determined
and the separation $R$ between the mass centers of the two parts is calculated,
with the shape having been divided according to the determined value
of the asymmetry $\alpha$.
A smooth function $R(q_2)$ is then obtained 
through a linear fit of these discrete values along the fusion valley.
Therefore, although the fusion path is one dimensional,
the separation coordinate $R(q_2)$ depends on all five shape parameters.

The equation of motion for the separation $R$ 
is calculated with the one-dimensional Langevin equation
\begin{equation}
\label{eq:langevin}
 \mu \ddot{R} = -\nabla_RU_I(R) - \gamma(R) \dot{R} - \sigma(R)\xi
\end{equation}
containing the conservative force exerted by the effective potential $U_I(R)$,
the friction force governed by the strength $\gamma$,
and the associated random force with strength $\sigma$.
The stochastic variable $\xi$ is drawn from a
standard normal distribution.
Furthermore,
 $\mu = M_PM_T/(M_P+M_T)$ is the reduced mass of the projectile and target.

While the friction strength $\gamma$ is negligible before contact,
it increases rapidly thereafter.
We approximate $\gamma$ by 
the radial component of the one-body window friction tensor \cite{blocki78:a}
which represents the dissipative effect of nucleon exchanges 
between the two collision partners,
\begin{equation}
\gamma = \mbox{$1\over2$} m_n \rho\overline{v}\,\pi c_{\rm neck}^2,
\end{equation}
where $m_n$ is the nucleon mass, $\rho$ is the nucleon number density
in nuclear matter, and
$\overline{v}=\frac{3}{4}v_{\rm Fermi}$ is the average nucleon speed.
Insertion of the standard values $m_n=939$\,MeV/$c^2$, 
$\rho=0.17\,{\rm fm}^{-3}$, and $v_{\rm Fermi}=0.27\,c$ yields
\begin{equation}
\gamma(R) \approx 16k\,\pi c_{\rm neck}^2\text{ MeV}/{\rm fm}\,c,
\label{eq:friction2}
\end{equation}
where the scaling parameter $k$ makes it possible 
to study the sensitivity of the results to the friction strength.
Unless otherwise indicated, the calculations
have been carried out for $k=1$.

As noted above, the neck radius increases quickly as the system moves past 
touching (see Fig.\ \ref{fig:neck}), so the friction strength $\gamma$,
being proportional to $c_{\rm neck}^2$, has a quite strong dependence on $R$.
For the cases considered, it quickly increases to about
$\gamma\approx500\text{ MeV}/{\rm fm}\,c$
after contact and,
as $q_2$ decreases further,
it continues to increase to about $\gamma\approx1200\text{ MeV}/{\rm fm}\,c$
at $q_2\approx1.9$
where $c_{\rm neck}\approx$ 4.9 fm.

At the separation $R$ the intrinsic excitation energy 
of the evolving system is given by
\begin{equation}
E^*(R)=E_{\rm cm}-U_I(R)-E_{\rm kin}(\dot{R}),
\end{equation}
where the kinetic energy associated with the radial motion is
$E_{\rm kin}(\dot{R})=\mbox{$1\over2$}\mu\dot{R}^2$
and the effective potential is $U_I(R)=U(R)+\hbar^2I(I\!+\!1)/2\mu R^2$.
The corresponding intrinsic temperature is given by
$T(R)=\sqrt{E^{\ast}(R)/a}$,
where $a=A_{\rm CN}/(8\text{ MeV})$ is the Fermi-gas level-density parameter.
In accordance with the fluctuation-dissipation theorem,
this quantity governs the magnitude of the random force in equilibrium
through $\sigma^2=2\gamma T$.

However, the finite relative motion enhances the fluctuations,
an effect that can be taken into account by means of
an effective temperature $T^*$ \cite{randrup79:a},
\begin{equation}
T^* = \left\langle\mbox{$1\over2$}\omega\coth{\omega\over2T}\right\rangle,
\end{equation}
where the exciton energy
$\omega=\dot{R}p_{\rm Fermi}$
is the typical amount of energy
dissipated in each nucleon transfer 
and the average is over all possible transfers.
To a good
approximation,
{\it i.e.} within 2\%,
$T^*$ can be obtained as
\begin{equation}
T^{\ast}(R)\approx \left[{m_n\over4\mu}E_{\rm F}E_{\rm kin}(\dot{R})
		   +T(R)^2\right]^{1\over2},
\end{equation}
where
$E_{\rm F}=p^2_{\rm Fermi}/2m_n\approx37$\,MeV
is the Fermi energy \cite{bohr69:a}.
The strength of the random force in Eq.\ (\ref{eq:langevin}) is then
\begin{equation}
\sigma(R)=\sqrt{2\gamma(R) T^*(R)}.
\end{equation}

\begin{figure}[bt]
\centering
\includegraphics[width=1.0\linewidth]{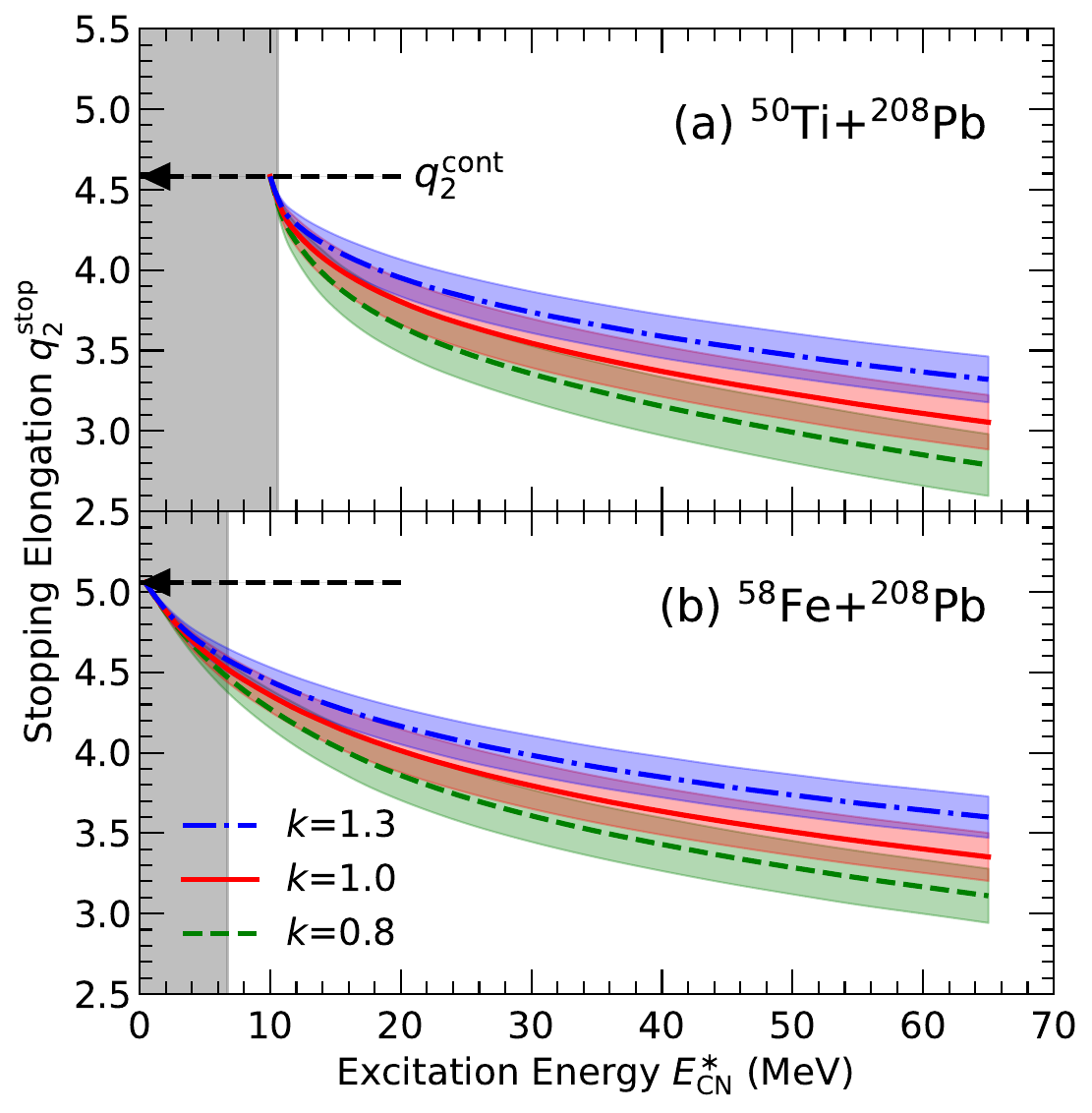}
\caption{\label{fig:q2_start}
Stopping elongation $q_2^{\rm stop}$ for the early drift-dominated dynamics
in $^{50}$Ti+$^{208}$Pb (a) and $^{58}$Fe+$^{208}$Pb (b)
using parameter $k=0.8$ (green), $k=1.0$ (red), or
$k=1.3$ (blue) in the friction coefficient in
Eq.\ (\ref{eq:friction2}).
The average values are at the center of the shaded bands
which have widths equal to the calculated dispersions 
of the distribution of $q_2^{\rm stop}$ for that excitation energy, 
$\sigma_{q_2}(E^{\ast}_{\rm CN})$.
The arrows mark the elongation $q_2^{\rm cont}$ of the contact configurations.
The gray regions correspond to tunneling energies.
}
\end{figure}

With the initial conditions 
\begin{equation}
R(0) = R^{\rm cont}, \qquad \dot{R}(0) = -\sqrt{2 E_{\rm kin}^{\rm cont}/\mu},
\end{equation}
where $E_{\rm kin}^{\rm cont}=E_{\rm cm}-U_I(R^{\rm cont})$
is the kinetic energy in the radial direction at contact,
the Langevin simulation along the fusion valley is continued 
until the radial speed has decreased to the mean equilibrium speed 
given by the equipartition theorem $\langle\dot{R}^2\rangle_{\rm eq}=T/\mu$.
The directed inward motion has then effectively stopped
and the diffusive evolution takes over (see
Sec.\ \ref{sec:method_diffusion}).
The corresponding elongation coordinate $q_2^{\rm stop}$ 
is obtained from the linear fit between $q_2$ and $R$ along the fusion valley.
The other four shape parameters are then determined by the fusion valley.

When the energy is increased, the inward motion proceeds further
and $q_2^{\rm stop}$ decreases.
Figure \ref{fig:q2_start} shows the stopping elongation
for the reactions $^{50}$Ti+$^{208}$Pb (a) and $^{58}$Fe+$^{208}$Pb (b) 
and for three different values of the friction scaling parameter $k$
in Eq.\ (\ref{eq:friction2}).
The average values are at the center of the shaded bands 
which have a width equal to the dispersion of the $q_2^{\rm stop}$ distribution 
for that energy, $\sigma_{q_2}(E^*_{\rm CN})$.
If no kinetic energy remains at contact, 
there is no inward drift and $q_2^{\rm stop}=q_2^{\rm cont}$
(marked with black arrows).
Because the present model does not describe tunneling, 
the energy has to be higher than the barrier 
for a reaction to occur.
This is indicated by the shaded gray region in Fig.\ \ref{fig:q2_start}.

While a higher initial energy thus results in 
more compact stopping configurations
(smaller $q_2^{\rm stop}$),
a stronger friction causes the inward motion to stop earlier,
thus resulting in a larger value of $q_2^{\rm stop}$.

\subsection{Diffusion dynamics}
\label{sec:method_diffusion}

From the stopping configuration $\boldsymbol{\chi}^{\rm stop}$, 
where the collective velocity has become negligible,
the system can diffuse in all five shape parameters.
In the limit of strong dissipation considered here,
the Langevin equation then reduces to the Smoluchowski equation
in which there is no kinetic energy and the various forces balance out \cite{abe96:a}.
The shape evolution then has the character of Brownian motion 
and can be simulated as a random walk 
on the multidimensional potential-energy surface \cite{randrup11:a}.
Such simulations are carried out here with the Metropolis method 
using the energy-dependent effective level density of Ref.\ \cite{randrup13:a}
that accounts for the gradual disappearance of the local
shell and pairing effects as the energy is raised.

A given walk is stopped and registered as a formation event 
if its elongation $q_2$ reaches that of the ground-state.
While the fission saddle shape at $q_2\approx1.5$ (see Fig.\ \ref{fig:epot1d})
is typically mass symmetric, 
the diffusive evolution will generally evolve toward compact shapes 
having mass asymmetries closer to the initial asymmetry.
However, once inside the saddle point,
the walks will diffuse towards the compound nuclear ground state 
and the evolution becomes quasiergodic.

More often, though, a walk may lead towards larger elongations
and an eventual division of the system into two fragments.
Such a QF event is assumed to occur if the neck radius
becomes smaller than 1.5\,fm \cite{albertsson21:a}.

\begin{figure}[bt]
\centering
\includegraphics[width=1.0\linewidth]{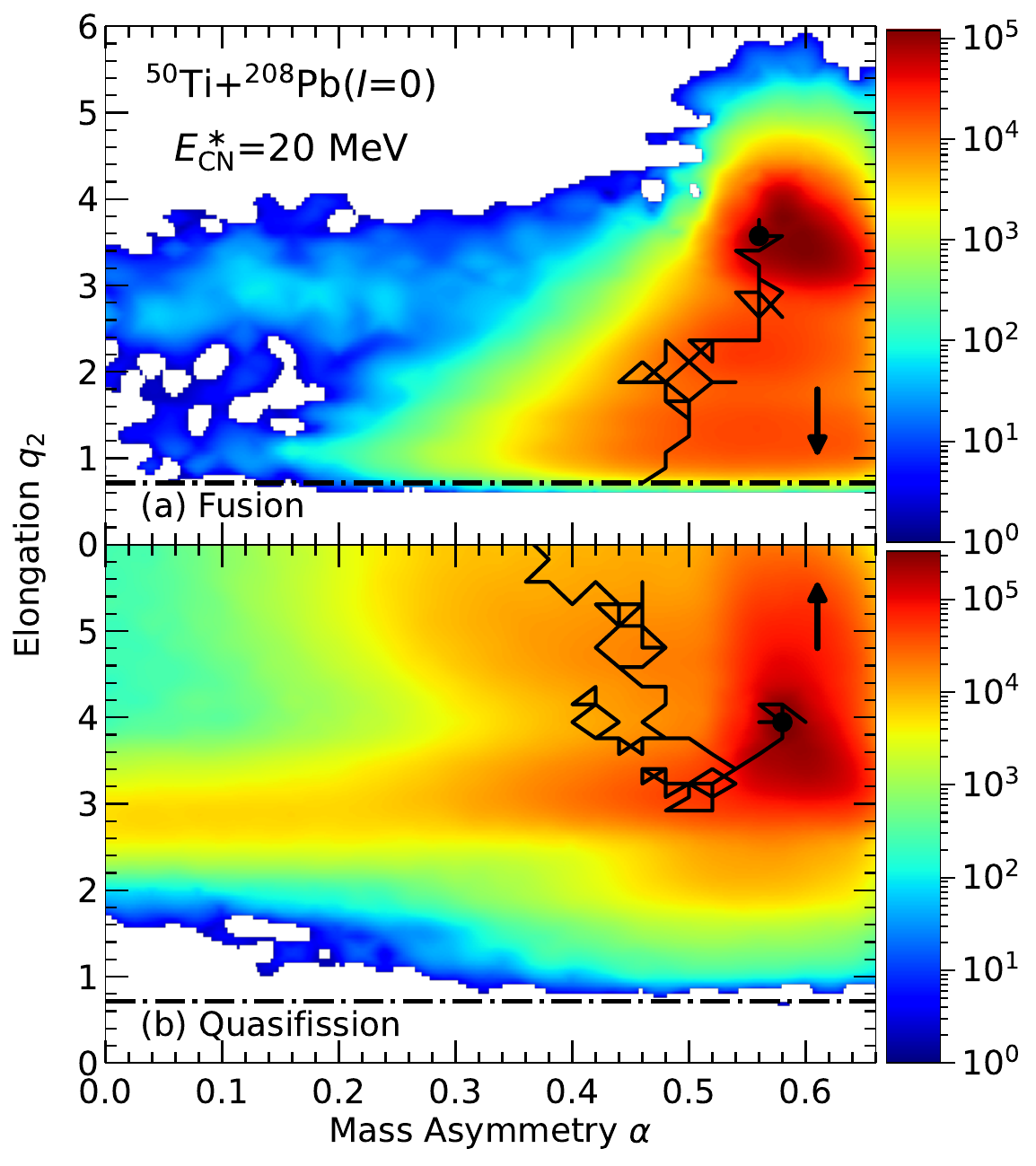}
\caption{\label{fig:visit_trajs}
Number of visits to shape lattice sites with a given
combination of asymmetry $\alpha$ and elongation $q_2$ for
the reaction $^{50}$Ti+$^{208}$Pb($I$=0)
with $E^\ast_{\rm CN}=20$ MeV.
(a) Walks resulting in fusion (28\%). (b) Walks resulting in QF (72\%).
In each panel, a typical trajectory (black path) is shown
with the black circle marking the starting point
and the arrow indicating the general direction of the shape evolution.
The ground-state elongation is shown as dotted-dashed line.
}
\end{figure}

Examples of walks are displayed in Fig.\ \ref{fig:visit_trajs} for the reaction $^{50}$Ti+$^{208}$Pb at $E^{\ast}_{\rm CN}=20$\,MeV and $I$=0.
The color contours show
the total number of visits to sites
with specified elongation $q_2$ and mass asymmetry $\alpha$
in Metropolis walks leading either to fusion (a) or to QF (b).
The respective probabilities for these event classes
are calculated to be 28\% and 72\%.
A typical trajectory is shown for each case,
with a black solid circle marking their starting point
(located at $(q_2,\alpha)\approx(3.8,0.57)$)
and the arrows indicating their general direction.
There is a large concentration of visits near the starting point
due to the presence of a minor local minimum in the fusion valley,
apparent in Fig.\ \ref{fig:epot1d}(b).
The fusion trajectories retain a relatively large mass asymmetry,
even for elongations $q_2 \approx 2.5$ where there is only a very low ridge
protecting the shape from diffusing into the fission valley.
However, after their  elongation $q_2$ has shrunk to that of the ground state,
the further diffusion leads the shape towards the actual ground-state shape
which has $\alpha\approx0$.
Those trajectories that do diffuse over the ridge and into the fission valley
have practically no chance of getting back over the fission barrier
towards more compact shapes and they therefore end up as QF events.

\section{Results}
\label{sec:results}

The dynamical calculations described above
are carried out for a range of angular momenta $I$.
For a given energy and a particular $I$
the formation probability
is calculated as the number of formation events
divided by the total number of events,
\begin{equation}
P_{\rm form}(E_{\rm CN}^{\ast},I)=\frac{\text{Number of formation events}}{\text{Total number of events}}.
\end{equation}
For each energy and angular momentum we simulate $10^5$ events.
For a given energy, reactions having a higher angular momentum $I$
will come to a stop further out in $q_2$,
because the effective potential energy $U_I(R)$
from the contact elongation and inwards grows ever steeper at larger $I$.
As a consequence, the subsequent diffusion process will be less likely
to lead to compact shapes and
the formation probability $P_{\rm form}(E^{\ast}_{\rm CN},I)$
will thus decrease as $I$ is increased.

\begin{figure}[bt]
\centering
\includegraphics[width=1.0\linewidth]{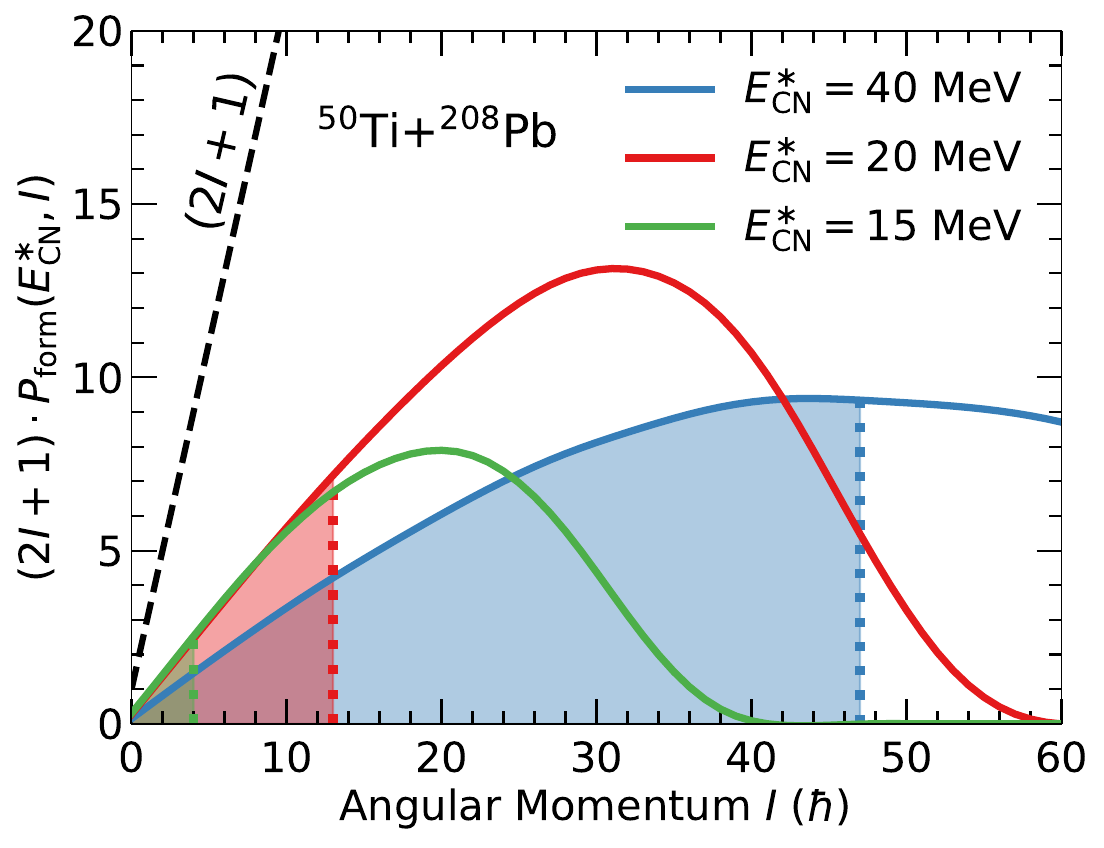}
\caption{\label{fig:formprob_iang}
$(2I+1)P_{\rm form}(E_{\rm CN}^{\ast},I)$ as a function of angular momentum $I$ versus the excitation energy $E_{\rm CN}^*$
in the reaction $^{50}$Ti+$^{208}$Pb.
Contributions below $I_{\rm max}$ are indicated by shaded regions.
}
\end{figure}

An {\it effective} formation probability can be obtained by
averaging over all angular momenta up to the maximum value $I_{\rm max}$
(see Fig.\ \ref{fig:Imax}), weighting each $I$ by the geometric factor $2I+1$,
\begin{equation}
\label{eq:pform_eff}
\langle P_{\rm form}(E_{\rm CN}^{\ast})\rangle =\frac{1}{(I_{\rm max}\!+\!1)^2}
\sum_{I=0}^{I_{\rm max}}(2I+1)P_{\rm form}(E_{\rm CN}^{\ast},I).
\end{equation}
Figure \ref{fig:formprob_iang} shows
$(2I\!+\!1)P_{\rm form}(E^{\ast}_{\rm CN},I)$
for the reaction $^{50}$Ti+$^{208}$Pb at various energies.
For each energy,
$E^{\ast}_{\rm CN}=15$, 20 and 40 MeV,
the contribution to $\langle P_{\rm form}(E_{\rm CN}^*)\rangle$
is shown by the
shaded region.
This
should be compared to the
corresponding area under the $2I+1$ line
which represents a 100\%\ formation probability.
The ratio between these two areas,
which thus constitutes the effective formation probability
in Eq.\ (\ref{eq:pform_eff}),
decreases with increasing energy,
with the values being
0.285, 0.270, 0.128,
respectively.
These values may be compared to the formation probabilites for $I=0$, which are found to be
0.286, 0.276, 0.162,
respectively.
Thus, the inclusion of angular
momentum implies a decrease in the formation probability of only
0.3\%, 2.2\%, and 21.0\%,
respectively.
This shows that in our calculations,
there is a much higher sensitivity to the excitation energy,
and a rather weak dependence on the angular momentum.

\begin{figure*}[bt]
\centering
\flushleft
\includegraphics[width=0.92\linewidth]{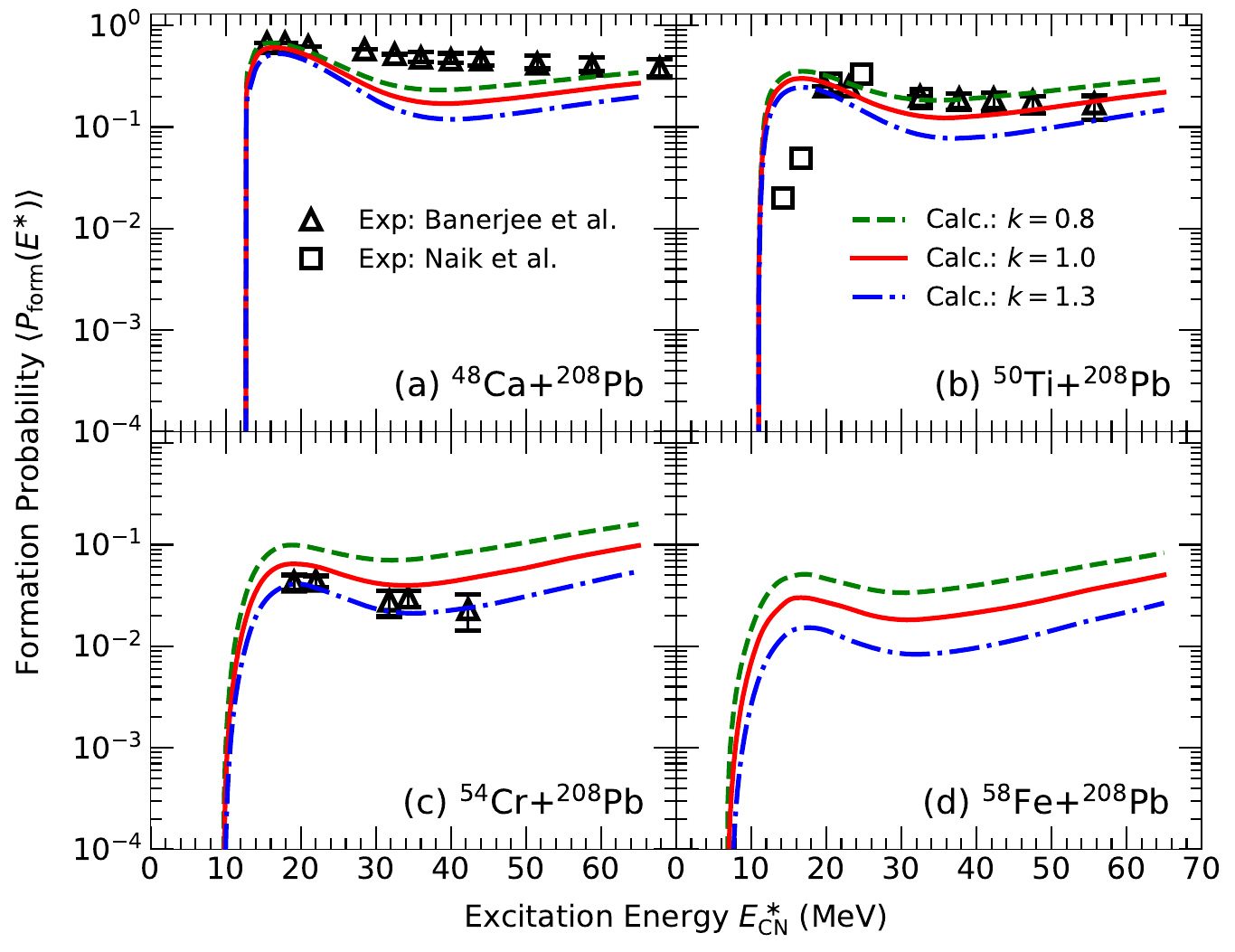}
\caption{\label{fig:form_tot}
Calculated effective formation probability
$\langle P_{\rm form}(E_{\rm CN}^*)\rangle$
as a function of the excitation energy $E_{\rm CN}^*$,
for three different dissipation strengths:
$k=0.8$ (green), 1.0 (red) and 1.3 (blue).
Data from Refs.\ \cite{banerjee19:a} and \cite{naik07:a} 
are shown as open black squares and triangles.
}
\end{figure*}

Figure \ref{fig:form_tot} shows the calculated effective formation probability
$\langle P_{\rm form}(E^{\ast}_{\rm CN})\rangle$ as a function of energy 
for the four considered reactions.
We note that the data of Naik {\it et al}.\ \cite{naik07:a} for $^{50}$Ti+$^{208}$Pb exhibit a maximum in the formation probability
at $E^{\ast}_{\rm CN}\approx25$ MeV.
The data of Banerjee {\it et al}.\ \cite{banerjee19:a},
which represent the upper limit of the formation probabilities,
instead show a steady decrease with
energy.

Because the model does not include tunneling,
formation is only possible if the total energy
 is sufficient to allow the system
to move up the fusion valley
and over the inner saddle.
The corresponding thresholds in $E_{\rm CN}^*$ are 
12.6, 10.6, 9.2, and 6.7 MeV for the four reactions, respectively.
For all four cases, the formation probability initially increases steeply 
above the threshold energy and then reaches a maximum 
at $E^*_{\rm CN}\approx15$\,MeV, followed by a moderate decrease.
This behaviour arises from the competition between two effects. 
As the energy is increased, 
the early inward motion stops at ever more compact shapes,
thereby increasing the probability that the subsequent diffusive evolution
succeeds in forming a compound nucleus.
However, it also becomes ever easier for the shape to find more favorable paths 
on the potential-energy landscape and cross the ridge from the fusion valley 
to the fission valley, leading to QF 
and thus decreasing the formation probability.
As already noted,
the angular momentum does not have a significant effect on these results.

\begin{figure}[bt]
\centering
\includegraphics[width=1.0\linewidth]{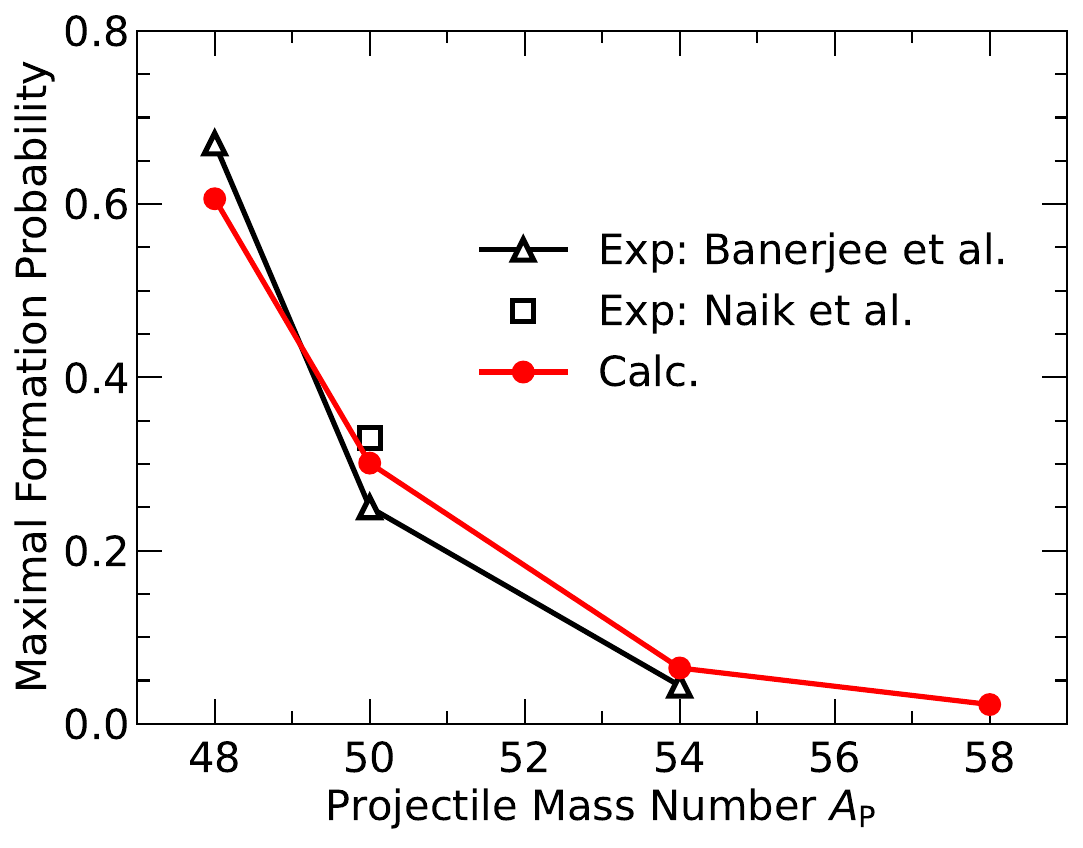}
\caption{\label{fig:form_max}
Maximum of the formation probability
as a function of the projectile mass number $A_{\rm P}$ 
in the four considered reactions.
The calculations are shown by red circles,
while the data (extracted from Fig.\ \ref{fig:form_tot})
are shown by open squares and triangles.
}
\end{figure}

In Fig.\ \ref{fig:form_tot} it is seen that the formation probability 
decreases as the projectile mass is increased.
The calculated {\it maximal} formation probability 
is shown in Fig.\ \ref{fig:form_max} versus the projectile mass,
together with the available data.
Ranging from about 60 percent for $^{48}$Ca to just a few percent for $^{58}$Fe,
the calculated values agree very well with the measured results.
The large formation probabilty for $^{48}$Ca is partly due to 
the contact configuration being relatively compact and partly due to
the slightly downward sloping potential energy in the fusion valley 
as one moves towards smaller values of $q_2$ (see Fig.\ \ref{fig:epot1d}a).
Both of these effects cause the early inward motion to proceed further,
so the diffusion process starts at more compact shapes 
and therefore needs fewer steps to form a compound nucleus.
For heavier projectiles, the contact configuration becomes more elongated 
and a gradual steeper upward slope in the fusion valley
(see
Fig.\ \ref{fig:epot1d})
leads to a steady reduction in the maximal formation probability.

At higher energies, from about 40 MeV and up,
the calculated formation probabilities increase steadily. 
This is due to the increasing temperature and the associated disappearance
of the shell and pairing effects which allows the diffusion process
to explore larger regions of the deformation parameter space, 
thereby also making it easier to attain more compact shapes, 
counted as fusion events.
However, the walks that do venture inside the inner saddle would 
relatively quickly cross the saddle again and proceed towards QF,
so an equilibrated compound nucleus is less likely to form.
Still, the comparison to the data of Ref.\ \cite{banerjee19:a} is meaningful, 
because the shapes in the registered formation events
have temporarily been very compact. 
During such a compact stage, the memory of the original direction of the symmetry axis will be lost and, upon separation, 
the angular distribution will then be that of a fusion-fission event. 
However, in experimental attempts to produce
SHN,
the increased formation probability at high energies 
will be of little importance because the very excited compound system
will promptly fission and the event will be registered as fusion-fission.

While the formation probability generally decreases when the friction
is stronger, the overall behavior of the curves in Fig.\ \ref{fig:form_tot}
is rather insensitive to the friction strength (governed by the parameter $k$),
The underestimate of the measured formation probability with the 
projectile $^{48}$Ca (a) could be remedied by reducing the friction strength.
Note, however, that the data points of Ref.\ \cite{banerjee19:a} are upper limits.
In turn,
the overestimate obtained with $^{54}$Cr (c) could be remedied
by using a slightly stronger friction strength.
Such adjustments of the friction strength 
are well within the current uncertainty on the nuclear dissipation.

\section{Summary and discussion}
\label{sec:summary}

We have calculated the probabilities
for compound-nucleus formation in several reactions with $^{208}$Pb
within the Langevin framework for the nuclear shape evolution.
After the colliding nuclei have come into contact,
the early evolution is dominated by an inward motion
during which initial radial kinetic energy is being dissipated.
The system then continues its shape evolution in a diffusive manner,
until it either reaches the region of compact shapes near the ground state
or redivides in a QF process.

The shape of the evolving system is described
in the three-quadratic-surfaces parametrization.
The associated 5D table of
the effective potential-energy surface 
includes shell and pairing effects that subside as the energy is raised.
During the early stage, the relative motion of the two parts of the system
is subject to the one-body window friction
while the subsequent diffusive shape evolution
is simulated as a random walk in the shape parameter space.
While no parameters were adjusted,
it was studied how the results depend on the dissipation strength.
Comparison with the experimental data suggests that
it might increase with increasing system size.

Relative to previous treatments of
SHN
formation
(see Refs.\ \cite{zagrebaev08:a,cap22:a}) several refinements have been made. 
The full dynamical process from contact to formation or separation is followed,
with the starting shape for the diffusive evolution 
being affected by energy-dependent nuclear structure effects 
obtained with well-established models \cite{moller09:a}.
The evolution of the system
is obtained by Monte Carlo event-by-event simulation,
which allows the extraction of a variety of correlation observables.

The calculated formation probabilities compare
very
well with available data,
as illustrated in
Figs.\  \ref{fig:form_tot} and \ref{fig:form_max}.
As the energy is increased, two opposite effects give rise to a maximum 
in the formation probability:
1) at higher energy the early inward evolution leads to more compact shapes
which increases the chance for the subsequent shape diffusion to reach
the compact region inside the fission barrier and thus form a compound nucleus;
2) the higher energy also makes a larger shape domain 
accessible to the diffusive evolution,
thereby facilitating the crossing of the ridge towards the fission valley,
hence decreasing the formation probability. 
At
even
higher energies shell and pairing effects gradually subside
and it becomes ever easier to enter the region of very compact shapes.
However, as the ground-state minimum then becomes less pronounced,
any compact shapes will quickly grow more elongated
and proceed to fission, thus being ineffective for
SHN
formation.

The gradual decrease in formation probability with increasing projectile mass 
(see Fig.\ \ref{fig:form_max}) was understood as being partly due to the fact
that heavier projectiles lead to more elongated contact configurations
and partly due to the changes in the slope of the potential energy 
along the fusion path (see Fig.\ \ref{fig:epot1d}).

Earlier studies with a diffusion-based model \cite{zagrebaev08:a}
show a steady increase of formation probability with energy.
It was therefore suggested in Ref.\ \cite{banerjee19:a} that 
diffusion is not the main mechanism that drives
SHN
formation
in fusion reactions with $^{208}$Pb nuclei.
This was addressed in the extended FBD model \cite{cap22:a} 
where the decrease in formation probability with increasing energy 
was attributed to suppression of contributions of higher partial waves 
in the reactions.
In contrast, the present results
display only a marginal effect of the angular momentum.
The decrease in formation probability with excitation energy seems instead 
to occur because more favorable QF paths become accessible
in the potential-energy landscape.

There have been indications that multi-nucleon transfer and energy dissipation
play a significant role already before contact is reached \cite{cook23:a}. 
This would result in a lower remaining kinetic energy at contact
and therefore more elongated shapes at the start of the diffusive stage.
The effect might be effectively emulated
by an increase in the friction strength.
Multi-nucleon transfer would also result in a distribution 
of $N/Z$ ratios in the two reaction partners at contact;
this degree of freedom is not
yet taken into account.

Generally, the formation of the compound system requires the
diffusive shape evolution to pass over the inner barrier.
However, at energies only slightly above the threshold energy,
shell and pairing effects in the saddle region 
may modulate the transmission probability 
because of the structure in the local density of states.
In particular, 
the absence of excited states between 0 and 2 quasiparticle states
is enhanced by the relatively large value of the pairing correlation 
in the barrier region, resulting in a non-monotonic energy dependence
of the formation probability.
This is
an effect similar to the undulating behavior of the symmetric fragment yield
in actinide fission, as discussed theoretically in Ref.\ \cite{Ward17}
and also observed experimentally \cite{Chadwick11,Glendening81}.
One might thus expect such structure effects to appear 
in the energy dependence of the formation probability 
just above the threshold energy.\\

\begin{acknowledgments}

The project was supported by the Knut and Alice Wallenberg foundation (KAW 2015.0021).
M.A.\ was supported by the Swedish Research Council under grant number 2022-00223 and J.R. was supported in part by the Office of Nuclear Physics
in the U.S.\ Department of Energy's Office of Science 
under Contract No.\ DE-AC02-05CH11231.
The computations were enabled by resources provided by the National Academic Infrastructure for Supercomputing in Sweden (NAISS) and the Swedish National Infrastructure for Computing (SNIC) at LUNARC.

\end{acknowledgments}

 \bibliographystyle{unsrt}


\end{document}